\documentstyle[sprocl]{article}
\def\be{\begin{equation}}
\def\ee{\end{equation}}
\def\bea{\begin{eqnarray}}
\def\eea{\end{eqnarray}}

\def\Rb {R_{\rm b}}
\begin{document}
\vspace*{-2cm}
\begin{tabbing}
\` DPF96 \\
\` 5a1/A05004 
\end{tabbing}
\vspace*{0.25cm}

\footnotetext{\it Presented at DPF96, American Physical Society Division
of Particles and Fields 1996 Divisional Meeting, Minneapolis, Minnesota,
10-15 August 1996.}

\title{A MEASUREMENT OF $\Rb$ USING MULTIPLE TAGS} 
\author{ ANDREW O. BAZARKO }
\address{CERN, CH-1211 Geneva 23, Switzerland \\
        representing the ALEPH COLLABORATION}
\maketitle\abstracts{
A measurement of $\Rb$ using five mutually exclusive hemisphere tags is 
presented. 
The preliminary result is 
$R_b = 0.2158  \pm 0.0009  (\mbox{stat.})\pm 0.0011  (\mbox{syst.})$. }

\section{Introduction}
Among the measurements available at the $Z^0$ pole, $\Rb$, 
the ratio of the $Z^0$ partial width into $b$ quarks and its total
hadronic partial width, is currently exciting particular interest.
Most electroweak and QCD 
radiative corrections cancel in the ratio, leaving $R_{\rm b}$ sensitive 
to corrections that couple preferentially to $b$ quarks,          
like the large CKM coupling to top quarks. As the parameters of 
the Standard Model become better constrained, a precise measurement of 
$\Rb$ tests the presence of novel vertex corrections.

This paper presents a new  measurement of $\Rb$ 
using data recorded by ALEPH in 1992-95, consisting of nearly 4 million
hadronic events. 
The measurement is 
performed using five hemisphere tags.\cite{5tags} 
The most important of
the tags is a lifetime-mass tag, 
offering high-purity $b$ tagging.\cite{Lifetime_paper}
Two additional $b$ tags are used, together with tags designed to 
select charm and $u,d$ and $s$ flavors. 
The latter permit the experimental determination
of the background efficiencies of the additional $b$ tags.
The five tags are mutually exclusive, so that a
hemisphere is tagged at most by one tag. 

\section{The Method}

Events are divided into hemispheres by the plane perpendicular to the
thrust axis.
The fraction of tagged hemispheres $f_{{\rm s},I}$ with tag $I$ is 
\begin{eqnarray}
  \label{singles}
  f_{{\rm s},I} = \Rb \epsilon_{{\rm b},I} +  R_{\rm c} \epsilon_{{\rm c},I} 
               + (1- \Rb - R_{\rm c}) \epsilon_{{\rm uds},I} & 
\end{eqnarray}
where 
$\epsilon_{a,I}$ are the hemisphere tagging efficiencies for 
flavor $a$ using tag $I$. 

The fraction of doubly tagged events $f_{{\rm d},IJ}$ 
with tags $I$ and $J$ is:
 \begin{eqnarray}
   \label{doubles}
   f_{{\rm d},IJ} = &\left[ R_{\rm b}\epsilon_{{\rm b},I}\epsilon_{{\rm b},J} 
                              (1+\rho_{{\rm b},IJ}) + 
               R_{\rm c}\epsilon_{{\rm c},I}\epsilon_{{\rm c},J} 
              (1+\rho_{{\rm c},IJ}) +\right.\nonumber \\ & 
              \left. (1- \Rb - R_{\rm c}) \epsilon_{{\rm uds},I}
                                          \epsilon_{{\rm uds},J} 
                  (1+\rho_{{\rm uds},IJ})\right] \: (2-\delta_{IJ})
 \end{eqnarray}
where $\rho_{a,IJ}$ are the hemisphere-hemisphere efficiency 
correlations for flavor $a$ and tags $I$ and $J$. 

Twenty quantities are measured:  5 single tag fractions 
and 15 double tag fractions.
Of the 62 unknown parameters in the above two equations,
$\Rb$ and 13 efficiencies are fitted to the data.
The two remaining efficiencies, $\epsilon_{c,Q}$ and $\epsilon_{x,Q}$
(the Q tag is described below),
and 
the 45 correlations 
are calculated using simulation. The systematic error reflects 
the uncertainties in these calculations.
$R_{\rm c}$ is taken as $0.171$ from electroweak theory.

\section{Event selection}

Using the data sample obtained on and near the $Z^0$ resonance with 
the ALEPH detector during the period 1992 to 1995, 
events must satisfy the following requirements:
 at least five reconstructed tracks;
 at least two jets clustered using the JADE algorithm \cite{jade} 
with energy greater than 10 GeV;
 $|\cos\theta_T|<0.65$, where $\theta_T$ is the angle between the beam and 
the thrust axis; and 
 $y_3<0.2$, where $y_3$ is the value of $y_{cut}$ that sets 
the transition from 2 to 3 jets.\cite{jade}
The remaining 2,059,066 events have a selection bias in 
favor of $b$ quarks relative to the lighter quarks of $0.23\pm 0.04$\% and 
a contamination of tau events of {$0.30\pm 0.01$\%.}

\section{The Five Hemisphere Tags}

In order to keep the hemisphere--hemisphere correlations small and
their calculation
transparent, the tag variables are constructed from quantities limited 
to the particular hemisphere only. A new primary vertex finder is
employed, which reconstructs the $Z^0$ decay point seperately in 
each hemisphere. 

The Q tag is a high purity $b$ tag 
formed from the combination of two tagging algorithms. 
Both algorithms make use of 3-dimensional track impact 
parameters, defined as the distance of closest approach of the track to the 
primary vertex.
The first algorithm provides the probability that all
of the hemisphere tracks originate from the primary vertex.\cite{aleph93} 
The second algorithm exploits the b/c hadron mass difference.
Tracks in a hemisphere are ordered inversely to their probability 
to originate from the primary vertex and are combined, in this
order, until the invariant mass of the combination exceeds 
1.8 $\rm GeV/c^2$. The probability 
of the last track added to have originated from the
primary vertex is used as a tagging variable.

The S tag uses the same lifetime-mass information 
as the Q tag but in a less pure $b$ tagging range, 
together with a neural net variable
that optimizes 25 event shape parameters for the selection of 
$Z^0$ decays to $b$ quarks.\cite{Henrard}

The L tag requires an identified muon or electron, with momentum greater 
than 3 GeV/c, and transverse
momentum with respect to its jet greater than 1.4 
GeV/c.\cite{Lepton_paper}

The X tag requires high probability that all tracks originate 
from the primary vertex, 
together with a requirement on the same neural net variable 
as the S tag, but in the opposite sense, in order to tag $uds$ events.

The C tag, which is the most difficult to construct, 
makes use of lifetime and event shape properties.
The tag uses a neural net variable that optimizes
one lifetime and 19 event shape parameters for the selection of 
$Z^0$ decays to $c$ quarks. In addition, tracks are divided into two groups
with rapidity with respect to their jets
greater or less than 5.1, chosen to get equal numbers of tracks
for $b$ hemispheres. The probabilities that the two groups of tracks
originate from the hemisphere primary vertex are used for tagging.

The Monte Carlo expectations for the $3\times 5$ efficiencies are shown 
in Table \ref{tab:MC_eff}. 

\begin{table*}[h]
\begin{center}
      \begin{tabular}{lccc}
\hline
        Tag \ & \ \ \ $\epsilon_{\rm uds}$ (\%) 
                 & \ \ \ $\epsilon_{\rm c}$ (\%)  
                 & \ \ \ $\epsilon_{\rm b}$  (\%) \\\hline
      Q   & $0.029$ &  $0.20$ & 19.55 \\ 
      S   & 0.200 &  1.40 & 17.57 \\ 
      L   & 0.156 &  0.69 &  4.25 \\
      X   & 11.70 &  3.96 &  0.23  \\ 
      C   & 7.93  & 16.20 &  2.59     \\\hline 
      \end{tabular}
\end{center}
    \caption{Tagging efficiencies as given by Monte Carlo simulation. }
    \label{tab:MC_eff}
\end{table*}

\section{Result}

The fit of $R_b$ and 13 of the 15 efficiencies
to the 20 measured tag fractions yields
\begin{equation}
\Rb = 0.21582 \pm 0.00087\nonumber
\end{equation}
with a $\chi^2$ of 8.1 for the 6 degrees of freedom, where the error is  
statistical. This result has been corrected for the event selection bias 
and tau contamination discussed in section 3.

\section{Systematic uncertainties}

The impact of a particular correlation on $\Rb$ is given by
  $\Delta \Rb / ( \Rb \: \Delta \rho_{a,IJ} ) $, i.e.
the relative uncertainty in $\Rb$ is the impact times the 
uncertainty in the correlation. The impacts vary a great deal, and 
uncertainties in only 
a dozen or so of the correlations contribute significantly 
to the uncertainty in $\Rb$. 
The correlations with the 
biggest impacts are $\rho_{{\rm b},QQ}$, with impact 0.45, and 
$\rho_{{\rm b},QS}$, with impact 0.42.
The impacts of the two Q tag background efficiencies are given by
\begin{displaymath}
\frac{\Delta \Rb}{\Rb} \; \frac{ \epsilon_{{\rm b},Q} }
                               { \Delta\epsilon_{{\rm c},Q} }= -1.5,
\; \; \; \; \; \; \; \;
\frac{\Delta \Rb}{\Rb} \; \frac{ \epsilon_{{\rm b},Q} }
                               { \Delta\epsilon_{{\rm uds},Q} }= -5.6
\end{displaymath}
A brief description of the sources of systematic uncertainty is
given below.

\subsection{Detector simulation uncertainty}

The Monte Carlo predictions for the Q tag background efficiencies depend
on the impact parameter resolution
and 
the efficiency for tracks to record a vertex detector hit. 
These quantites are measured in the data and remaining 
uncertainties 
produce corresponding errors in $\epsilon_{{\rm c},Q}$ and 
$\epsilon_{{\rm uds},Q}$.

\subsection{Systematics from $b$ and $c$ physics uncertainties} 

Uncertainties in the physics
charmed and light flavored hadrons propagate dominantly through the
Q tag background 
efficiencies, whereas uncertainties in bottom physics 
propagate through the correlations. 
To assess these uncertainties
the physics inputs to the Monte Carlo simulation are varied within 
their allowed experimental ranges.\cite{lepewwg} 
The largest contribution found with this procedure is due to the 
rate of gluon splitting into heavy quarks.

\subsection{Hemisphere-hemisphere correlation uncertainties}

The correlations have three origins: detector inhomogeneities,
the size and position of the interaction region, 
and the coupling of the $B$ hadron momenta through 
gluon radiation and hadronization. 
The Monte Carlo simulation's ability to reliably calculate 
these sources of correlation is checked against data by studying
the correlation due to variables linked to these correlating effects.
Uncertainties are assigned that are in addition to those due to $b$ and $c$
physics that have been considered above.

The correlation contributions in the following four variables
are compared between data and simulation:
the cosine of the angle between the thrust and
the beam axes, the component of the error in the reconstructed 
hemisphere primary 
vertex transverse to the thrust axis, the momenta of the two jets (because 
of their
correlation with the $B$ hadron momenta), and $y_3$ (to check the 
gluon radiation-induced correlations). 
The larger of either a measured difference between data and simulation or
the precision of the comparison is assigned as an uncertainty in 
the correlation.
The uncertainty in $\Rb$  
is found by propagating and adding their contributions linearly, 
so as to take correlations between the 
correlations into account. 

A summary of the systematic errors is given in Table \ref{tab:sysunc}.

\begin{table}[h]
  \begin{center}
    \leavevmode
    \begin{tabular}{cc}
\hline
 source                              &  $\Delta R_b$ \\\hline
 detector simulation                 &  0.00050 \\
 Monte Carlo statistics              &  0.00047  \\
 event selection                     &  0.00010  \\
 physics uncertainties               &  0.00083  \\
 hemisphere correlations             &  0.00033  \\
\hline
    \end{tabular}
    \caption{ Systematic uncertainties.}
    \label{tab:sysunc}
  \end{center}
\end{table}

\section{Conclusions}

The fraction of hadronic $Z^0$ decays to $b$ quarks, $R_b$, has been measured 
using five mutually exclusive tags. The preliminary value is:

\begin{equation}
  R_b = 0.2158 \pm 0.0009 (\mbox{stat.}) \pm 0.0011 (\mbox{syst.}) 
                  -0.019\times (R_{\rm c}-0.171)  \nonumber
\end{equation}
where the first error is statistical, the second is systematic, and 
the explicit dependence on $R_{\rm c}$ is given. 

This measurement includes the 1992 data used in the earlier ALEPH $\Rb$
measurement with a lifetime tag\cite{aleph93} and 
therefore supercedes it.


\end{document}